\documentclass[10pt,notitlepage,a4paper,aps,prd,onecolumn,superscriptaddress,nofootinbib,groupedaddress]{revtex4-1}
\usepackage{array}

\usepackage[dvipsnames]{xcolor}

\usepackage[pdfencoding=auto, psdextra]{hyperref}
\usepackage[T1]{fontenc}
\usepackage[utf8]{inputenc}
\usepackage[english]{babel}
\usepackage{hyperref}
\usepackage[mathscr]{euscript}
\usepackage{caption}
\usepackage{subcaption}
\hypersetup{
colorlinks = true,
linkcolor = blue,
filecolor = magenta,
urlcolor = blue,
pdftitle = {TLSs interaction},
pdfpagemode = FullScreen,
citecolor = blue}
\usepackage{physics}

\usepackage{caption}

\usepackage{indentfirst}

\usepackage[pdftex]{graphicx}

\usepackage{amsmath}
\allowdisplaybreaks  % para que pueda partir fórmulas que ocupan más de una línea, necesita el paquete anterior
\usepackage{amssymb} % para cargar algunos símbolos como \blacksquare y \square
\usepackage{amsfonts} % para cargar algunas fuentes en estilo matemático
\usepackage{enumerate}

\usepackage{fancyhdr}                     % para definir distintos tipos de cabeceras y pies de página

\usepackage{float}                             

\usepackage{amsmath}         % For advanced math typesetting
\usepackage{amsfonts}        % For additional mathematical fonts
\usepackage{amssymb}         % For more math symbols
\usepackage{amsthm}          % For theorem environments
\usepackage{empheq}          % For enhanced equation environments

\usepackage{color}           % For colored text
\usepackage{ragged2e}        % For ragged right alignment
\usepackage{titlesec}        % For customizing section title formats

\usepackage{tipa}            % For the thorn symbol
\usepackage[utf8]{inputenc} % For UTF-8 input encoding
\allowdisplaybreaks          % Allow page breaks within equations

\usepackage{enumitem}        % For customizable lists

\usepackage{tikz}            % For creating graphics
\usetikzlibrary{shapes.geometric,arrows.meta,arrows,positioning}  % Load specific TikZ libraries
\usetikzlibrary{tikzmark}   % For advanced TikZ marking

\titleformat{\paragraph}
  {\normalfont\normalsize\bfseries}{\theparagraph}{1em}{} % Format paragraph titles
\titlespacing*{\paragraph}
  {0pt}{3.25ex plus 1ex minus .2ex}{1.5ex plus .2ex} % Set spacing for paragraph titles

\DeclareUnicodeCharacter{03B2}{\ensuremath{\beta}}

\begin{document}
\title{On the role of true and false chirality in producing parity violating energy differences}

\author{Daniel Martínez-Gil}
\email{daniel.martinez@ua.es}
\affiliation{Fundacion Humanismo y Ciencia, Guzmán el Bueno, 66, 28015 Madrid, Spain.}
\affiliation{Departamento de F\'{\i}sica Aplicada, Universidad de Alicante, Campus de San Vicente del Raspeig, E-03690 Alicante, Spain.}

\author{Pedro Bargueño}
\email{pedro.bargueno@ua.es}
\affiliation{Departamento de F\'{\i}sica Aplicada, Universidad de Alicante, Campus de San Vicente del Raspeig, E-03690 Alicante, Spain.}

\author{Salvador Miret-Artés}
\email{s.miret@iff.csic.es}
\affiliation{Instituto de Física Fundamental, Consejo Superior de Investigaciones Científicas, Serrano 123, 28006, Madrid, Spain}

%%%% Abstract text to be placed here %%%%%%%%%%%%
\begin{abstract}
In this work we tackle the problem of showing which type of influences can lift the degeneracy between truly and falsely chiral systems, showing that
only when both systems and influences are both truly (falsely) chiral, a parity violating energy difference between left- and right-handed systems can be produced.  In particular, after considering the enantiomers of a chiral molecule as paradigmatic truly chiral systems, we rigorously show, 
under a quantum field theoretically approach, that only a truly chiral influence such as the $Z^{0}$-mediated electroweak interaction can lift the degeneracy between enantiomers. On the contrary, we explicitly show that a falsely chiral influence, such as an axion-mediated interaction in chiral molecules, can not 
lift the aforementioned degeneracy. These results extend Barron's seminal ideas [L. D. Barron, True and false chirality and parity violation, Chem. Phys. Lett
{\bf 123}, 423 (1986)] to a quantum field theory-based approach.
\end{abstract}
%%%%%%%%%%%%%%%%%%%%%%%%%%%

\maketitle
%%%%%%%%%% Insert the texts which can accomdate on firstpage in the tag "fmtext" %%%%%

\section{Introduction}
Living organisms are composed of a system of organized molecules with a specific chirality. Ribonucleic acid and deoxyribonucleic acid (RNA and DNA), responsible for genetic information replication and storage, are made of D-sugars (R-type molecule) \cite{Schrodinger}, resulting in the most stable configurations for DNA and RNA being right-handed helices. It turns out that all sugars found in nature are D-sugars, and the 20 existing amino acids are of the L-type. This is called molecular homochirality  as, in principle, there should be an equal probability of molecules with left-handed or right-handed chirality, but in nature, only one type is found. This peculiar fact is one of science's most intriguing and fundamental problems, and its solution is yet to be determined \cite{Jones}.
\\
\\
The hypothesis that life emerged from a racemic environment and that homochirality developed later had many followers before 1990. In 1987, it was argued that the origin of chiral molecules on Earth "must have occurred at the same time as the origin of life or shortly after" \cite{BADA198721}. It should be emphasized, however, that there is no experimental evidence to support such speculations. Nevertheless, both theoretical and experimental data indicate that "life could not have existed without molecular asymmetry" \cite{terent1957origin}. We live in a world with DNA and RNA as the basis of life, whose structure is determined by the chirality of the molecules that compose them, so all evidence leads to the same conclusion: without chirality, there is no life.
\\
\\
Throughout history, many mechanisms have been proposed to produce molecular homochirality, such as circularly polarized light \cite{braun}, magnetic fields \cite{rikken}, electrons in beta decays \cite{ulri}, or mechanisms based on parity violation \cite{temperatura, radiacion, autocatalitico, Yamagata, Salam, Pedro3,Bargueno2009}. Some years ago, an excess of L-amino acids was discovered in meteorites \cite{10,11}. This finding reinforced the idea of an extraterrestrial origin of biological homochirality \cite{WABONNER}. In this context, some mechanisms, such as parity violation in weak interactions, would acquire special interest \cite{meteoritosMacDermott}.
\\
\\
In the present work we will focus on the latter, specifically on the idea of the existence of a parity violating energy difference (PVED) between L and R molecules. We want to clarify that when we write "PVED", we refer to the parity violating energy difference produced by any parity-violating interaction. Although this PVED has not yet been detected, differences between vibrational frequencies of chiral molecules due to it, ($\Delta \nu ^{PV}$), are currently being tested at the Laboratoire de Physique des Lasers in Paris. This group achieved 
 a sensitivity of $\frac{\Delta \nu ^{PV}}{\nu} \approx 10^{-13}$ \cite{Paris1999}.  Even more, it is expected that, for specific molecules composed of Ruthenium (Ru) and Osmium (Os) atoms, the theoretical value for that measurement is $\frac{\Delta \nu ^{PV}}{\nu} \approx 10^{-14}$ \cite{experimentos}. Therefore, we think we are close to observe the PVED in chiral molecules in the near future. It is also remarkable that due to this parity-violating effect, a particular enantiomer becomes more stable and more likely to exist, producing an enantiomeric excess.  In order to produce a homochiral state it is necessary to first produce an enantiomeric excess (due to PVED in our case) and finally to amplify it. The intriguing hypothesis that PVED might serve as the origin of the enantiomeric excess ultimately amplified into the asymmetry observed in life was first proposed by Ulbricht in 1959 \cite{Ulbricht1, Ulbricht2} and later by Yamagata in 1966 \cite{Yamagata}. During the 1980s, Kondepudi and Nelson \cite{Kondepudi} developed stochastic models based on a Frank-type autocatalytic network \cite{FRANK1953459}, enabling them to explore the sensitivity of the SMSB (Spontaneous Mirror Symmetry Breaking) process to extremely weak chiral influences. Their calculations suggested that the energy values required to bias the SMSB process toward a single direction were within the range of PVED values estimated for biomolecules. Despite the competing influence of random fluctuations, as highlighted by Lente in subsequent analyses \cite{Lante}, it remains plausible that this low "asymmetric factor" could guide the system toward a preferred asymmetric state with significant probability \cite{Kondepudi2} (the interested reader can see a recent review of biological homochirality in \cite{RevewHomochirality}). 
\\
\\
In order to classify fields and forces capable of producing enantiomeric excess in chiral molecules, Barron introduced two fundamental concepts: true and false chirality \cite{barronfund}, which will be explained in the following section. Barron showed that only a truly chiral influence can induce PVED in chiral molecules \cite{barronpv}.
 His results were based on a direct application of the properties of both quantum states representing chiral molecules and interaction Hamiltonians under $\hat{P}$, $\hat{T}$, and spatial rotations. In particular, Barron proposed as quintessential true and false chiral influences the weak neutral currents and axions, respectively \cite{Barron07}. In fact, all his derivations were performed under a non-relativistic formalism for both the systems under study (chiral molecules) and interaction Hamiltonians for example weak neutral currents and axions). In this manuscript, it is our purpose to generalize Barron's results in two different ways: (i) by analyzing the possibility of producing PVED in a truly (falsely) chiral system under a truly (falsely) chiral influence and (ii) by treating both system (chiral molecules) and influences (weak neutral currents and axions) under a quantum field theory (QFT)-based approach.
\\
\\
Specifically, this work is organized as follows: In Sec. \ref{sec2} we introduce the concepts of true and false chirality, which are essential to the rest of the manuscript. In Sec. \ref{sec3} we show that only a truly chiral influence can produce PVED in chiral molecules, extending Barron's conclusions \cite{barronpv} showing that also a falsely chiral influence acting on a falsely chiral system can induce PVED.
In Sec. \ref{sec4} we give a QFT-based proof that the PVED in chiral molecules can only be produced by a truly chiral influence, exemplified by considering
a fully relativistic electroweak influence within these molecules. In addition, we also show that a falsely chiral interaction can not lift the degeneracy between the enantiomers of a chiral molecule, as a particular kind of axion-molecule interaction exemplifies. Finally, Sec. \ref{sec5} provides the conclusions of the present work.

\section{True and False Chirality}\label{sec2}
\begin{figure}[ht]
\centering
\includegraphics[width= 5cm]{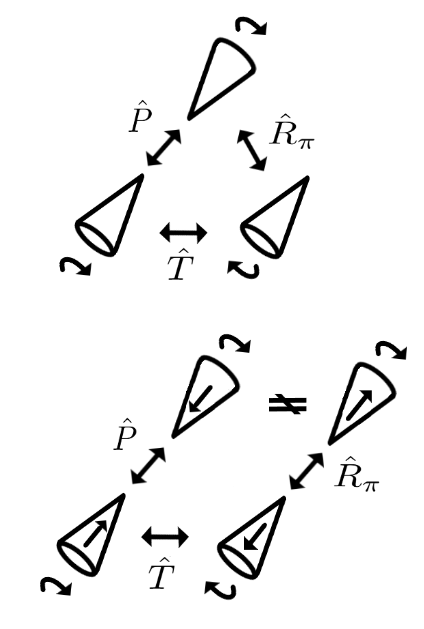}
\caption{\justifying In the top part of the figure, the effects of parity ($\hat{P}$), time inversion ($\hat{T}$), and a $\pi$ rotation ($\hat{R}_\pi$) on a rotating cone are shown. The bottom part of the figure shows the effect of the same symmetries when applied to a rotating and translating cone. Adapted from \cite{barron1}.}
\label{conos}
\end{figure}

\begin{figure}[ht]
\centering
\includegraphics[width= 7cm]{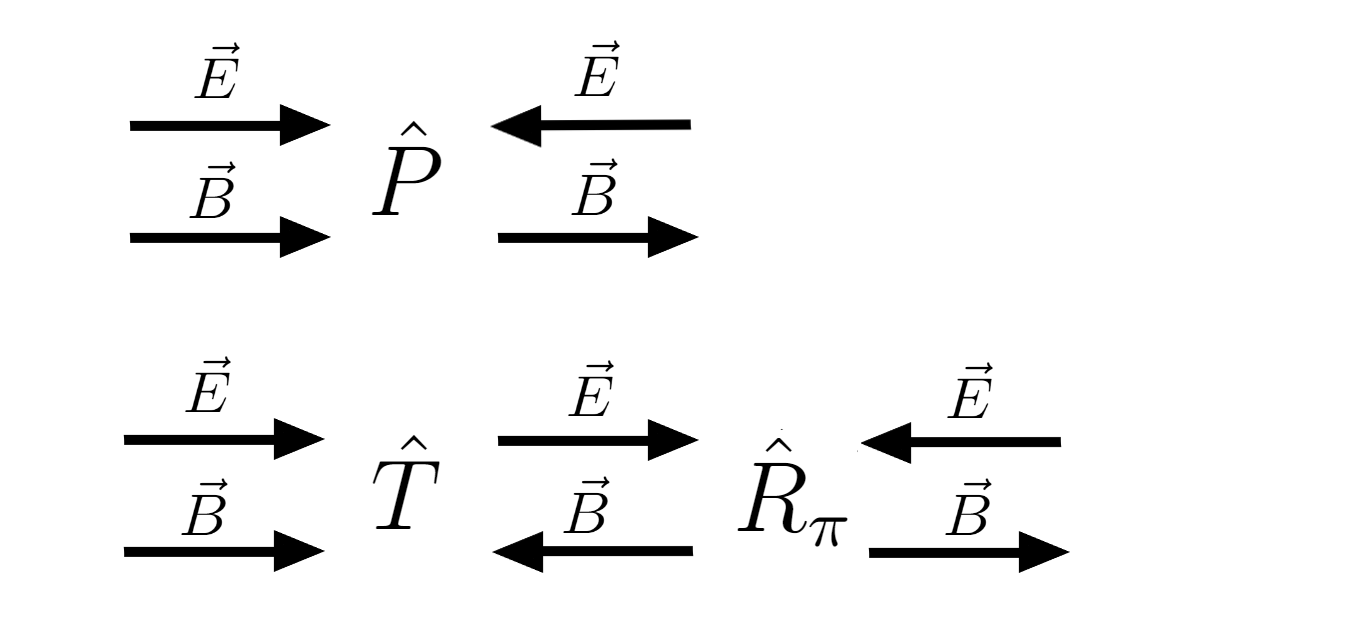}
\caption{\justifying Chirality test applied to a system with parallel electric and magnetic fields.}
\label{eb}
\end{figure}

We all agree when the term {\it chiral} applies to stationary objects, meaning that the resulting image is not superimposable with the original object when you apply a reflection (or parity operation) to it. However, when this definition is applied to abstract objects or physical systems, the concept of chirality becomes less clear.
\\
\\
The definition of true chirality was first mentioned by Barron in 1981 \cite{barronfund}, but it was not until 1986 when the same author gave a precise definition of it \cite{barronpv}. In his definition of true chirality, he states \cite{barronpv}, "true chirality is possessed by systems that exist in two distinct enantiomeric states that are interconverted by space inversion but not by time reversal combined with any proper spatial rotation". On the contrary, false chirality occurs in systems that exist in two different enantiomeric states, which are interconverted by spatial inversion as well as time inversion followed by spatial rotation. These definitions can be better understood with the help of Fig. (\ref{conos}). The top part of the figure corresponds to a falsely chiral system since applying $\hat P$ to the rotating cone is equivalent to applying $\hat T$ followed by a $\pi$ rotation. The bottom part shows a clear example of a truly chiral system, since applying the aforementioned symmetries results in different outcomes. From now on, we will refer to the ``chirality test" as the process of applying these three symmetries and then checking if the outcomes match.
\\
\\
Regarding physical entities, both electric and magnetic uniform fields are commonly used as interesting examples. A uniform electric field $\Vec{E}$ is a time-even polar vector, and the magnetic field $\Vec{B}$ is a time-odd axial vector. Let us consider a system with parallel electric and magnetic fields; 
%\textcolor{red}{Dice que aquí especifiquemos que nos referimos a una sola molécula, pero estoy hablando de un caso general de un campo eléctrico y magnético, no es que pertenezca a ninguna molécula};  
if we apply the chirality test, we can see that this system is falsely chiral, as applying $\hat{P}$ is equivalent to applying $\hat{T}$ followed by $\hat R_{\pi}$ (see Fig. (\ref{eb})).
\\
\\
Therefore, if we want a truly chiral combination as a result of the multiplication of two fields, one of them must be a time-odd axial vector, and the other one must be a time-odd polar vector, resulting in a time-even pseudoscalar, violating parity while respecting time and rotational invariance. 
\\
\\
At this point, let us comment on some misunderstandings in the context of true and false chirality. As was previously mentioned, a system being either truly or falsely chiral must exist in two distinct enantiomeric states. It is commonly believed that this fact implies a parity violation in the system, which is not true. A distinction between (spontaneous) symmetry breaking and symmetry violation must be done \cite{libroBarron}. On one hand, symmetry breaking occurs when the system displays a lower symmetry than its underlying intrinsic/internal Hamiltonian. Thus, neglecting the small parity-violating term in the Hamiltonian, the symmetry operation that the Hamiltonian possesses but the chiral molecule lacks is parity, and it is the parity operation that interconverts the two enantiomeric parity-broken states \footnote{We acknowledge one of the anonymous referees
for this comment.}. Another example of parity-broken states is circularly polarized light (CPL). Although CPL exists in two different enantiomeric states, its intrinsic Hamiltonian is purely electromagnetic, respecting the parity symmetry. Therefore, CPL has two parity-broken (not parity-violating) states.
On the other hand, symmetry violation refers to the lack of symmetry in the Hamiltonian, so the associated law of conservation of symmetry no longer holds. For example, if the parity operator and the Hamiltonian do not commute, the Hamiltonian has a parity-violating term (which is the case of weak neutral current interactions).

\subsection{Systems and influences}
During the 1850s, Pasteur realized that molecular homochirality on Earth must have been induced by a universal chiral force or a natural influence. In this sense, not only systems (like the ones mentioned earlier) but also influences have to be included in our description. From now on we will differentiate between truly and falsely chiral systems and influences. We refer to the object under study as "the system" which, in the present work, are generally chiral molecules, and we refer to "the influence" as the agent (interaction) capable of inducing a specific chirality in the considered system.
%In the literature, `asymmetric synthesis' refers to cases where there has been no external influence for the existence of enantiomeric excess, and `absolute asymmetric synthesis' refers to cases where such an influence has been present. Here, we specifically focus on the absolute asymmetric synthesis of a chiral molecule induced by an influence, whether truly or falsely chiral.
\\
\\
At this point, some comments are in order. The influence we are referring to is encoded in a specific interaction Hamiltonian, which does not need to undergo the chirality test to determine if it is truly chiral or not. The Hamiltonian must violate parity (otherwise, it would not be chiral). For it to be truly chiral, it must respect both time inversion and rotation; otherwise, it will be considered falsely chiral.
\\
\\
From these previous considerations, we can see that there are four different cases for the system-influence combination, considering that both the influence and the system can be truly or falsely chiral. In the following section, we will show which of these four combinations gives place to a non-zero PVED. 
\\
\\
To avoid misunderstandings, we want to clarify here what we mean as influence. To do this, we consider the interaction between CPL and a chiral molecule. It is known that both CPL and chiral molecules are truly chiral. In this case, it is usual to think that CPL is the influence (being a truly chiral influence), but it is not in our terminology. We call influence to the interaction Hamiltonian between CPL and the chiral molecule, which is not truly chiral because it is purely electromagnetic (and it does not violate parity). %Although CPL can not produce a difference of energy between the L and R enantiomers (because the interacting Hamiltonian respects parity), it can produce optical activity in chiral molecules, which is a truly chiral observable \cite{Barron01}.}
\\
\\
\section{Existence of PVED and system-influence chirality}\label{sec3}
As a specific case of a truly chiral system, we use chiral molecules, which are of interest in this work. We consider that: (i) the parity-converted left-handed chiral molecule is a right-handed one, and vice versa and (ii) both left and right-handed states are invariant under time inversion. (iii) It also makes sense to think that a chiral molecule remains the same molecule after rotation. 
\\
\\
Therefore, the conditions for a state to represent a single chiral molecule are:
\begin{align}\label{tqcond}
\hat{P}\psi_L =& \psi_R. &\hat{T}\psi_L =& \psi_L , & \hat{R}\psi_L =& \psi_L,
\end{align}
where $\psi$ represents a state of the system (the molecule).
\\
\\
The condition for a falsely chiral system is
\begin{equation}\label{circled2}
\hat{P}\psi = \hat{R}_\pi \hat{T}\psi.
\end{equation}
\\
\\
Regarding influences, we have three conditions for a truly chiral influence:
\begin{equation*}
\hat{P}H\hat{P}^{-1} = -H,
\end{equation*}
\begin{equation}\label{circled3}
\hat{T}H\hat{T}^{-1} = H^\dagger,
\end{equation}
\begin{equation*}
\hat{R}\pi H\hat{R}\pi^{-1} = H,
\end{equation*}
and for a falsely chiral influence, we use the following properties:
\begin{equation*}
\hat{P}H\hat{P}^{-1} = -H,
\end{equation*}
\begin{equation}\label{circled4}
\hat{T}H\hat{T}^{-1} = -H^\dagger,
\end{equation}
\begin{equation*}
\hat{R}\pi H\hat{R}\pi^{-1} = H.
\end{equation*}
\\
\\
With these definitions in place, we can now proceed with the four cases of the system-influence combination.
\\
\\
 \textbf{i) Truly chiral system and truly chiral influence}
\\
\\
     In this case, we apply the chirality test to the system-influence combination, using conditions \eqref{tqcond} and \eqref{circled3}. First, apply $\hat{P}$ to an initial state, and then $\hat{R}_\pi$ and $\hat{T}$ to that initial state and compare the results. We denote the truly chiral Hamiltonian as $H_{TC}$.
\\
\\
On one hand, we have

\begin{equation*}
    \bra{\psi_L} H_{TC} \ket{\psi_L} = \bra{\hat{P}\psi_R} H_{TC} \ket{\hat{P}\psi_R} = 
\end{equation*}
\begin{equation}\label{pexpression}
    = \bra{\psi_R} \hat{P}^\dagger H_{TC} \hat{P}\ket{\psi_R} =  -\bra{\psi_R} H_{TC} \ket{\psi_R}.
\end{equation}
\\
\\
On the other hand,
\begin{equation*}
    \bra{\psi_L} H_{TC} \ket{\psi_L}  = \bra{\hat{T}\psi_L} H_{TC} \ket{\hat{T}\psi_L} =
\end{equation*}
\begin{equation*}
    = \left(\bra{\psi_L} \hat{T} ^\dagger H_{TC} \hat{T}\ket{\psi_L}\right)^\dagger = \left(\bra{\psi_L} H_{TC}^\dagger \ket{\psi_L}\right)^\dagger =
\end{equation*}
\begin{equation*}
    = \bra{\psi_L} H_{TC} \ket{\psi_L} = \bra{\hat{R}_\pi\psi_L} H_{TC} \ket{\hat{R}_\pi\psi_L} = 
\end{equation*}
\begin{equation}\label{ec7}
    = \bra{\psi_L} \hat{R}_\pi ^\dagger H_{TC} \hat{R}_\pi\ket{\psi_L} = \bra{\psi_L} H_{TC}\ket{\psi_L}.
\end{equation}
\\
\\
Therefore, after applying both $\hat{T}$ and $\hat{R}_\pi$  operators, no new information is obtained. From the application of the $\hat{P}$ operator, we obtain that the degeneracy between the L and R enantiomers is broken, meaning that it could exist an energy difference between both enantiomers. In this case,
this energy difference between enantiomers \footnote{In fact, $\epsilon_{PV}$ is the difference between the expected value of the energy of the L and R enantiomers, but in the literature is often referred to as {\it energy difference} due to an abuse of language.}, defined as
\begin{equation}
\epsilon_{PV} = \frac{\bra{\psi_L} H\ket{\psi_L} - \bra{\psi_R} H\ket{\psi_R}}{2} = \frac{H_{LL}-H_{RR}}{2},
\end{equation}
can be different from 0. 
\\
\\
\textbf{ii) Truly chiral system and falsely chiral influence}
\\
\\
We consider now conditions \eqref{tqcond} and \eqref{circled4}. The Hamiltonian is now falsely chiral, and it is denoted as $H_{FC}$. Let us note that the
behaviour of both $H_{TC}$ and $H_{FC}$ under $\hat P$ is the same. Therefore, considering the $\hat{T}$ and $\hat{R}_\pi$ parts, we have
\begin{equation*}
    \bra{\psi_L} H_{FC} \ket{\psi_L}  = \bra{\hat{T}\psi_L} H_{FC} \ket{\hat{T}\psi_L} = 
\end{equation*}
\begin{equation*}
    =  - \bra{\psi_L} H_{FC} \ket{\psi_L} = -\bra{\hat{R}_\pi\psi_L} H_{FC} \ket{\hat{R}_\pi\psi_L} = 
\end{equation*}
\begin{equation}
    = -\bra{\psi_L} H_{FC}\ket{\psi_L}.
\end{equation}
\\
\\
Using the result from \eqref{pexpression}, we arrive at this triple equality
\begin{equation*}
    H_{LL} = -H_{LL} = -H_{RR},
\end{equation*}
concluding that $\epsilon_{PV} = 0$.
%However, please note that, at first sight , it may seem like the main value of the energy of both enantiomers is 0, but this is not the case since the reference point is set at the potential minimum of the molecule. It is the difference of the main value of the energies which is equal to 0. For simplicity, we refer to this as energy difference, not mentioning the main value.
\\
\\
\textbf{iii) Falsely chiral system and truly chiral influence}
\\
\\
Now, we consider conditions \eqref{circled2} and \eqref{circled3}. The procedure for falsely chiral systems changes slightly because now we must satisfy condition \eqref{circled2}. Following this condition, as $\hat{P}$ exchanges the L and R enantiomers, $\hat{R}_\pi\hat{T}$ will do the same. We must remember that this system is falsely chiral, so it no longer represents a chiral molecule (it does not meet conditions \eqref{tqcond}). With these comments at hand, we have
\begin{equation*}
    \bra{\psi_L} H_{TC} \ket{\psi_L}  = \bra{\hat{R}_\pi\hat{T}\psi_R} H_{TC} \ket{\hat{R}_\pi\hat{T}\psi_R} = 
\end{equation*}
\begin{equation}
\label{eqnew}
    = \left(\bra{\psi_R} \hat{T} ^\dagger \hat{R}_\pi^\dagger H_{TC} \hat{R}_\pi\hat{T}\ket{\psi_R}\right)^\dagger =\bra{\psi_R}H_{TC}\ket{\psi_R}.
\end{equation}
\\
\\
Combining \eqref{eqnew} and \eqref{pexpression}, we arrive at
\begin{equation*}
    H_{LL} = H_{RR} = -H_{RR}.
\end{equation*}
Therefore, as in the previous case, we conclude that $\epsilon_{PV} = 0$.
\\
\\
\textbf{iv) Falsely chiral system and falsely chiral influence}
\\
\\
In this case, we apply conditions \eqref{circled2} and \eqref{circled4} to obtain
\begin{equation*}
    \bra{\psi_L} H_{FC} \ket{\psi_L}  = \bra{\hat{R}_\pi\hat{T}\psi_R} H_{FC} \ket{\hat{R}_\pi\hat{T}\psi_R} = 
\end{equation*}
\begin{equation}
    = -\bra{\psi_R}H_{FC}\ket{\psi_R}.
\end{equation}
\\
\\
Therefore, also in this case a PVED can be induced.
\\
\\
Therefore, we conclude that the effect of an influence on the system depends on their chiral behavior, as the following table summarizes: 
\\
\\
\begin{center}
\begin{tabular}{|c|c|c|}
\hline
\textbf{System} & \textbf{Influence} & \textbf{Conclusions} \\
\hline
\hline
$TC$ & $TC$ & $H_{LL} = -H_{RR}$ \\
\hline
$TC$ & $FC$ & $H_{LL} = H_{RR} = 0$ \\
\hline
$FC$ & $TC$ & $H_{LL} = H_{RR} = 0$ \\
\hline
$FC$ & $FC$& $H_{LL} = -H_{RR}$ \\
\hline
\end{tabular}
\end{center}
\vspace{0.5cm}
Let us note that the only option for PVED to occur is when both the system and the influence have the same chiral behavior. However, when dealing with chiral molecules (which are the central theme of this work), we remark that only a truly chiral influence can produce PVED. In addition, we note that 
$H_{LL} = 0$ or $H_{RR} = 0$ does not mean that the energy of the corresponding enantiomers is vanishing. In fact, since the total Hamiltonian can be expressed as
\begin{equation}
    \hat{H} = \hat{H}_0 + \hat{H}_{PV},
\end{equation}
where $\hat{H}_{PV} = \hat{H}_{TC}$ or $\hat{H}_{PV} = \hat{H}_{FC}$ depending on the case that we are working on, $H_{LL} = H_{RR}  = 0$ means that the energy of the enantiomers is not changed by the parity violating effect that we are considering.

\subsection{Examples of truly and falsely chiral influences}

An example of a truly chiral influence is the interaction between electrons and nucleons, mediated by a $Z^0$ boson (see, for example, \cite{BERGER2004188} and references therein)
\begin{equation}\label{Hamiltonianoew}
H_{NC} = \frac{G_F}{4\sqrt{2}}\sum_i\sum_j Q_{W_i} \{ \Vec{p}_j \cdot \Vec{\sigma}_j,\rho(\Vec{r_i}-\Vec{r}_j)\},
\end{equation}
where $G_F$ is the Fermi constant, $\Vec{p}$ is the momentum, $\Vec{\sigma}$ is the spin of the electrons, $\rho$ is the nucleon density (electrons and nucleons are indexed by $j$ and $i$, respectively). Additionally, $Q_W$ is given by
\begin{equation}
Q_W = (1-4\sin^2\theta_W) Z-N,
\end{equation}
where $\theta_W$ is the Weinberg angle and $Z$ and $N$ represent the number of electrons (or protons) and neutrons, respectively. It is worth noting that the Hamiltonian \eqref{Hamiltonianoew} can be obtained under a non-relativistic approximation within QFT \cite{BERGER2004188}.
\\
\\
It can be easily verified that this Hamiltonian represents a truly chiral influence. This behavior is due to the scalar product of $\Vec{\sigma}_e$ (time-odd axial vector) and $\Vec{p}$ (time-odd polar vector), so it transforms like a time-even pseudoscalar, violating $\hat{P}$ but respecting $\hat{T}$. As a result, this Hamiltonian can produce PVED in chiral molecules \cite{Barron2012}.
\\
\\
On the other hand, the same electron-nucleon interaction mediated by an axion represents a falsely chiral influence. In natural units ($c = \hbar = 1$), the interaction Hamiltonian \cite{moody} would have the following form
\begin{equation}\label{axion}
H_{ax} = (g_s^N g_p^e) \frac{\Vec{\sigma}_e \cdot \Vec{r}}{8\pi m_e} \left(\frac{m_{\phi}}{r}+\frac{1}{r^2}\right) e^{-m_\phi r},
\end{equation}
where $g_s^N$ is the scalar coupling constant of the axion to a nucleon, and $g_p^e$ is the pseudoscalar coupling constant to the electron, $\Vec{r}$ represents the separation vector between the electron and the nucleon, and $m_\phi$ is the mass of the axion.
\\
\\
In the Hamiltonian \eqref{axion}, all terms are scalars except for $\Vec{\sigma}_e$ (time-odd axial vector) and $\Vec{r}$ (time-even polar vector). Therefore, it transforms as a time-odd pseudoscalar, violating both $\hat{P}$ and $\hat{T}$. Thus, it represents a falsely chiral influence (and, therefore, it cannot produce PVED in chiral molecules) \cite{Barron2012}.
\\
\\
Finally, let us mention that, although enantiomers affected by a falsely chiral influence are strictly degenerate, an enantiomeric excess can be produced by the very same falsely chiral influence \cite{barron1,Barron02}. This fact can occur when the system is out of thermodynamic equilibrium, permitting differences in the rate constants. In essence, when the non-linearity of the autocatalytic process is large enough, it leads to a high entropy production. Therefore, 
instability of the non-equilibrium racemic state can occur and a new chiral attractor ({\it i. e.}, a macroscopic set of molecules) may appear, which leads to the formation of macroscopic chiral states \cite{PCCP2022}. An example of this effect was recently shown in \cite{Micali} (see also \cite{barronnature}). 
\\
\\
At this point, we would like to emphasize again that the only option for PVED to occur is when both the system and the influence have the same chiral behaviour. Importantly, regarding this chiral behaviour, we have assumed certain transformation rules of both systems and influences under the $\hat R_{\pi}$, $\hat P$ and $\hat T$ operations. 
%But the validity of these rules can only be rigorously applied within the machinery of QFT. 
For instance, we remind the reader that we have considered that (i) the parity-converted left-handed chiral molecule is a right-handed one, and vice versa; (ii) both left and right-handed states are invariant under time inversion and (iii) it makes sense to think that a chiral molecule remains the same molecule after rotating it. Even more, \eqref{Hamiltonianoew} and \eqref{axion} are, as commented before, nonrelativistic approximations of fully 
quantum relativistic processes. Therefore, it would be desirable to extend the previous results by describing both systems and influences within QFT. This is the objective of the rest of the manuscript.

\section{System-influence chirality in quantum field theory}\label{sec4}

In this section, we are focused on the role of parity violation in chiral molecules described within QFT. The basic idea is to realize that the
electroweak interaction couples molecular electrons to nuclei through a $Z^{0}$ boson. In this sense, as it can be inferred from \eqref{Hamiltonianoew}, it is the electron helicity (often termed as electron chirality \cite{Hegstrom1991}) that couples to the nuclear density. Therefore, considering that the coupling of this electron chirality with the nucleon density is the responsible for the parity-violating part of the interaction, and given that electrons and nucleons are described by bispinors within QFT, here we assume that our truly chiral systems (chiral molecules) can be represented by bispinors (in fact, since the implementation of a parity-odd (time-even) perturbation operator into a fully relativistic four-component Dirac-Hartree-Fock framework treatment of the parity violating energy differences in chiral molecules \cite{Laerdahl1999}, these type of techniques have been employed to different levels of sophistication). In this sense, only two possible situations, depending on the chiral type of the influence, will be possible. As examples of truly and falsely chiral influences, we use the electron-nucleon interaction mediated by weak neutral currents for the former and mediated by axions, for the latter, as mentioned previously. We begin with the electroweak case. 

\subsection{Weak neutral currents as truly chiral influences}

We first consider the electron-nucleon interaction mediated by the $Z^0$ boson which, consistently with its non-relativistic approximation, should be a truly chiral influence, meaning that it should satisfy the conditions \eqref{circled3}.
\\
\\
By using standard Feynman rules, the Hamiltonian for the electron-nucleon interaction through a weak neutral current is given by 
 \begin{equation}\label{HEW}
     H_{NC} = \frac{G_F}{\sqrt{2}}\sum_i\sum_j:\Bar{N}_i(t,\Vec{x}) \gamma^\mu (V^N - A^N \gamma^5)N_i(t,\Vec{x}) \Bar{e}_j(t,\Vec{x}) \gamma_\mu (V^e - A^e \gamma_5)e_j(t,\Vec{x}):,
 \end{equation}
where the notation $: A :$ refers to the normal ordering of the $A$ operators, $N(t,\Vec{x})$ and $e(t,\vec{x})$ are the nucleon and electron fields, respectively, $V^N$,  $A^N$, $V^e$ and $A^e$ are coupling constants, $\gamma^\mu$ are the Dirac matrices satisfying 
$\{\gamma^{\mu},\gamma^{\nu}\}= 2 \eta^{\mu\nu} $, where $\eta^{\mu\nu}$ is the Minkowskian metric with signature $(-,+,+,+)$ and the index $i$ and $j$ refers to a sumation over nucleons and electrons. We also use the notation $\Bar{\psi} \equiv \psi^\dagger \gamma^0$ for the Dirac adjoint. 
\\
\\
The previous Hamiltonian can be regrouped into four different terms according to their coupling constants as
  \begin{equation}
     H_{NC} = H_{VV} -H_{AV}-H_{VA} +H_{AA},
 \end{equation}
\\
\\
where $V$ and $A$ refer to vector and axial vector, respectively.
\\
\\
As we are only interested in the parity violating terms of the Hamiltonian, we note that these terms can be expressed as
\begin{equation}
   H_{AV} =  \Bar{N}(t,\Vec{x}) \gamma^\mu A^N \gamma^5 N(t,\Vec{x}) \Bar{e}(t,\Vec{x}) \gamma_\mu V^e e(t,\Vec{x}),
\end{equation}
\begin{equation}
    H_{VA} = \Bar{N}(t,\Vec{x}) \gamma^\mu V^N N(t,\Vec{x}) \Bar{e}(t,\Vec{x}) \gamma_\mu A^e \gamma_5 e(t,\Vec{x}).
\end{equation}
\\
\\
For our purposes,  $H_{AV}$ and $H_{VA}$ are equivalent, so we will only use $H_{AV}$ along the rest of the manuscript. 
\\
\\
At this point, some comments are in order.  In the Weyl basis, bispinors can be expressed as a combination of left-handed (L) and right-handed (R) chiral spinors, in the form
\begin{equation}\label{Weyl}
    \psi = \begin{pmatrix}
\psi_L \\
\psi_R
\end{pmatrix},
\end{equation}
where
\begin{align}
    P_L\psi =& \begin{pmatrix}
\psi_L \\
0
\end{pmatrix} \equiv \psi_L^0,& P_L =& \frac{1-\gamma^5}{2},
\end{align}
\begin{align}
    P_R\psi =& \begin{pmatrix}
0 \\
\psi_R
\end{pmatrix} \equiv \psi_R^0,& P_R =& \frac{1+\gamma^5}{2},
\end{align}

where $P_{L,R}$ are the so-called chirality projectors.
\\
\\
According to \cite{peskin}, bispinors behave as follows under a parity transformation: 
\begin{equation}
    \hat{P} \psi(t,\Vec{x}) \hat{P} = \eta_a \gamma^0 \psi (t,-\Vec{x}),
\end{equation}
\\
where $\eta_a$ are possible phases. Therefore, a parity-transformed $\psi_L^0$ can be expressed as
\begin{equation}\label{quiralLR}
    \hat{P}\psi_L^0 (t,\Vec{x}) = \eta_ a P_L \gamma^0 \psi (t,-\Vec{x}) = \eta_a \psi_R^0 (t,-\Vec{x}).
\end{equation}
\\
\\
Once these previous considerations have been taken into account, we can now proceed with the case of weak neutral currents as truly chiral influences. Let us start with the first part of the chirality test: behaviour under parity.

\subsubsection{Behaviour under parity}
Using the following transformation rules \cite{peskin}
\begin{equation*}
    \hat{P}\Bar{\psi}\gamma^\mu \psi\hat{P} =  \begin{cases} +\Bar{\psi} (t,-\Vec{x})\gamma^\mu \psi (t,-\Vec{x}) & \text{for }  \mu = 0, \\-\Bar{\psi} (t,-\Vec{x})\gamma^\mu \psi (t,-\Vec{x}) & \text{for } \mu = 1,2,3, \end{cases}
\end{equation*}
\begin{equation*}
    \hat{P}\Bar{\psi}\gamma^\mu \gamma^5\psi\hat{P} =  \begin{cases} -\Bar{\psi} (t,-\Vec{x})\gamma^\mu \gamma^5\psi (t,-\Vec{x}) & \text{for } \mu = 0, \\+\Bar{\psi} (t,-\Vec{x})\gamma^\mu \gamma^5\psi (t,-\Vec{x}) & \text{for } \mu = 1,2,3, \end{cases}
\end{equation*}
\\
\\
we end up with 
\\
\\
    \begin{equation*}
    \bra{\psi_L^0 (t,\Vec{x})} H_{AV} (t,\Vec{x})\ket{\psi_L^0 (t,\Vec{x})} =\bra{\hat{P}\psi_R^0 (t,-\Vec{x})} H_{AV} (t,\Vec{x})\ket{\hat{P}\psi_R^0 (t,-\Vec{x})} =
\end{equation*}
\begin{equation*}
    =
    \bra{\psi_R^0 (t,-\Vec{x})} \hat{P}\Bar{N}(t,\Vec{x}) \gamma^\mu A^N \gamma^5 N(t,\Vec{x}) \Bar{e}(t,\Vec{x}) \gamma_\mu V^e e(t,\Vec{x})\hat{P}\ket{\psi_R^0 (t,-\Vec{x})} =
\end{equation*}
\begin{equation*}
=-\bra{\psi_R^0 (t,-\Vec{x})}  H_{AV} (t,-\Vec{x})\ket{\psi_R^0 (t,-\Vec{x})}.
\end{equation*}

Therefore, an overall minus sign is obtained in this process, as it should be. But, in addition, an extra minus sign inside the coordinates of the Hamiltonian and bispinors appears. We will return to this extra sign at the end of the section.

Let us now proceed with the other part of the chirality test: behaviour under time reversal and rotation.

\subsubsection{Behaviour under time reversal}

Although in the first part of the manuscript we were restricted by using the L and R states, we do not have this restriction while using bispinors. Therefore, in order to extend the previous results to the QFT formalism, we have to find a bispinor $\phi$ which satisfies
\begin{equation}
\hat{T}\phi(-t,\Vec{x}) = \psi_L^0 (t,\Vec{x}).
\end{equation}
\\
\\
In order to find $\phi$, we must use the transformation law of bispinors under time inversion \cite{peskin},
\begin{equation}
\hat{T}\psi(t,\Vec{x})\hat{T} = (\gamma^1\gamma^3) \psi(-t,\Vec{x}),
\end{equation}
which means that $\phi$ fulfills
\begin{equation}\label{T}
\hat{T}\phi(t,\Vec{x}) = (\gamma^1\gamma^3) \phi(-t,\Vec{x}) = \psi_L^0(-t,x).
\end{equation}
\\
\\
Therefore, if we express $\psi_L^0$ as
\begin{equation}
    \psi_L^0 = \begin{pmatrix}
 \psi_{L1}\\
\psi_{L2} \\
0\\
0 
\end{pmatrix},
\end{equation}
\\
and substitute it in the Eq. \eqref{T}, we find that
\begin{equation}
    \phi (-t,\Vec{x}) = \begin{pmatrix}
-\psi_{L2}(-t,\Vec{x}) \\
\psi_{L1}(-t,\Vec{x}) \\
0\\
0
\end{pmatrix}.
\end{equation}
\\
\\
Let us note that the L (or R) bispinor is not invariant under time inversion, which would seem to go against the condition \eqref{tqcond} stating that any enantiomer is invariant under $\hat{T}$. However, this last consideration can be taken to be only an approximation outside QFT. 
\\
\\
After taken into account the antiunitary character of $\hat T$ and using  \cite{peskin}
\\
\begin{equation*}\label{T23}
    \hat{T}\Bar{\psi}\gamma^\mu \gamma^5\psi =  \begin{cases} -\Bar{\psi} (-t,\Vec{x})\gamma^\mu \gamma^5\psi (-t,\Vec{x})\hat{T} & \text{for } \mu = 0, \\+\Bar{\psi} (-t,\Vec{x})\gamma^\mu \gamma^5\psi (-t,\Vec{x})\hat{T} & \text{for } \mu = 1,2,3\end{cases}
\end{equation*}
\begin{equation*}\label{T11}
    \hat{T}\Bar{\psi}\gamma^\mu \psi\hat{T} =  \begin{cases} +\Bar{\psi} (-t,\Vec{x})\gamma^\mu \psi (-t,\Vec{x}) & \text{for } \mu = 0, \\-\Bar{\psi} (-t,\Vec{x})\gamma^\mu \psi (-t,\Vec{x}) & \text{for } \mu = 1,2,3 \end{cases}
\end{equation*}
\\
\\
we arrive to

\begin{equation*}
\bra{\psi_L^0 (t,\Vec{x})} H_{AV} (t,\Vec{x})\ket{\psi_L^0 (t,\Vec{x})} = \bra{\hat{T}\phi(-t,\Vec{x})} H_{AV} (t,\Vec{x})\ket{\hat{T}\phi (-t,\Vec{x})} =
\end{equation*}
\begin{equation*}
= -\left( \bra{\phi(-t,\Vec{x})}\hat{T}\Bar{N}(t,\Vec{x}) \gamma^\mu A^N \gamma^5 N(t,\Vec{x}) \Bar{e}(t,\Vec{x}) \gamma_\mu V^e e(t,\Vec{x})\hat{T}\ket{\phi (-t,\Vec{x})} \right)^\dagger =
\end{equation*}
\begin{equation*}
= +\left( \bra{\phi(-t,\Vec{x})}\Bar{N}(-t,\Vec{x}) \gamma^\mu A^N \gamma^5 N(-t,\Vec{x}) \Bar{e}(-t,\Vec{x}) \gamma_\mu V^e e(-t,\Vec{x}) \ket{\phi (-t,\Vec{x})} \right)^\dagger =
\end{equation*}
\begin{equation*}
= + \bra{\phi(-t,\Vec{x})}\left( N^\dagger(-t,\Vec{x}) \gamma^0 \gamma^\mu A^N \gamma^5 N(-t,\Vec{x}) e^\dagger(-t,\Vec{x}) \gamma_0 \gamma_\mu V^e e(-t,\Vec{x}) \right)^\dagger\ket{\phi (-t,\Vec{x})} =
\end{equation*}
\begin{equation*}
= + \bra{\phi(-t,\Vec{x})} \Bar{e}(-t,\Vec{x}) V^e \gamma_\mu e(-t,\Vec{x}) \Bar{N}(-t,\Vec{x}) A^N \gamma^{\mu} \gamma^{5} N(-t,\Vec{x}) \ket{\phi (-t,\Vec{x})} =
\end{equation*}
\begin{equation}\label{ultimaT}
= + \bra{\phi(-t,\Vec{x})} H_{AV} (t,\Vec{x}) \ket{\phi (-t,\Vec{x})}.
\end{equation}

Let us now go with the rotation part of the chirality test.

\subsubsection{Behaviour under rotations}

A spinor rotation can be represented by the rotation operator 
\begin{equation*}
\hat{R}_\theta = \exp{-i\Vec{\theta} \cdot \frac{\Vec{\sigma}}{2}} = \exp{-i\abs{\theta}\hat{n}\cdot \frac{\Vec{\sigma}}{2}} = 
\end{equation*}
\begin{equation}
=\cos\left(\frac{\abs{\theta}}{2}\right)-i \hat{n}\cdot \Vec{\sigma}\sin\left(\frac{\abs{\theta}}{2}\right),
\end{equation}
where the second equality can be obtained by expanding the exponential in a Taylor series and rearranging terms according to the Taylor expansions of sine and cosine functions. The vector $\Vec{\theta}$ represents the rotation angle, $\Vec{\sigma}$ is a vector whose components are the three Pauli matrices and $\hat{n}$ represents the unitary rotation axis. 
\\
\\
To be consistent with the definition of true chirality, we must perform a rotation of $\pi$ radians, which can be represented by
\begin{equation}\label{rotacionsigma}
    \hat{R}_\pi = -i\hat{n}\cdot \Vec{\sigma} = \begin{cases} -i\sigma_1 & \text{for } \hat{n} = \begin{pmatrix}
        1\\
        0\\
        0
    \end{pmatrix}, \\-i\sigma_2 & \text{for } \hat{n} = \begin{pmatrix}
        0\\
        1\\
        0
    \end{pmatrix}, \\
    -i\sigma_3 & \text{for } \hat{n} = \begin{pmatrix}
        0\\
        0\\
        1
    \end{pmatrix}.\end{cases}
\end{equation}
\\
\\
Although we have defined spinor rotations, we remind the reader that we are working with bispinors, which are not an irreducible representation and can be expressed in terms of two spinors as $(\frac{1}{2},0) \oplus (0, \frac{1}{2})$. This allows us to express the rotation operator for bispinors as
\begin{equation}\label{rotacion}
    \hat{R}_{\theta}^{bi} = \begin{pmatrix}
        \hat{R}_\theta & 0\\
        0 & \hat{R}_\theta
    \end{pmatrix}.
\end{equation}
\\
\\
In addition, it is important to note that in QFT we are not restricted to L and R states, this fact permitting the existence of a third possibility after applying the rotation operator. Therefore, depending on the rotation that we apply, three cases are found:
\\
\\
\textbf{i) The system is neither a L or a R bispinor}
\\
\\
This is the case when 
\begin{equation}
    \phi(x,t) = \hat{R}_{\pi}^{bi} \eta (x,t)    ,
\end{equation}
where $\eta$ is different from $\psi_L^0$ and $\psi_R^0$. This case is not interesting because it does not permit us to complete the chirality test, which
can only be done by comparing the $L$ and $R$ states of the system (in this case we would reach something as $H_{LL} = H_{\eta\eta}$, which is an irrelevant condition for our purposes).
\\
\\
\textbf{ii) The system is a R bispinor}
\\
\\
This is the case when
\begin{equation}
    \phi(x,t) = \hat{R}_{\pi}^{bi} \psi_R^0 (x,t).
\end{equation}
\\
A comment on this case should be made. By looking at the expression \eqref{rotacion}, the rotation matrix is purely diagonal in blocks, which means that if we apply a rotation to the bispinor $\psi_R^0$ we can never recover the $\phi$ bispinor. Therefore, this case can not be achieved.
\\
\\
 \textbf{iii) The system is a L bispinor}
\\
\\
This is the case when
\begin{equation}\label{case3rotation}
    \phi(x,t) = \hat{R}_{\pi}^{bi} \psi_L (x,t).
\end{equation}
\\
\\
In order to see if this is possible, we have to find a $\pi$ rotation that changes the state $\psi_L^0$ into the state $\phi$. If we express the Pauli matrices as
\begin{align}
    \sigma_1 =& \begin{pmatrix}
        0&1\\
        1&0
    \end{pmatrix}, &         \sigma_2 =& \begin{pmatrix}
        0&-i\\
        i&0
    \end{pmatrix}, &         \sigma_3 =& \begin{pmatrix}
        1&0\\
        0&-1
    \end{pmatrix}, 
\end{align}
we see that the only combination which transform $\psi_L^0$ into $\phi$ is a $\pi$ rotation around the $y$ axis. Following Eq. \eqref{rotacionsigma}, the corresponding rotation matrix would be
\begin{equation}\label{rotacion2}
    \hat{R}_{\pi_{2}}^{bi} = \begin{pmatrix}
        0 & -1 & 0 &0\\
         1 & 0 & 0 &0\\
          0 & 0 & 0 &-1\\
           0 & 0 & 1 &0
    \end{pmatrix}.
\end{equation}
\\
\\
\\
Therefore, from these previous considerations, only case (iii) will be considered in order to complete the chirality test by introducing the appropriate rotation operator given by \eqref{rotacion2}. 
\\
\\
From \eqref{ultimaT}, we obtain
\begin{equation*}
\bra{\phi(-t,\Vec{x})} H_{AV} (-t, \Vec{x}) \ket{\phi (-t,\Vec{x})}=
\end{equation*}
\begin{equation*}
    = \bra{\hat{R}_{\pi_{2}}^{bi} \psi_L^0(-t,\Vec{x})} H_{AV} (-t, \Vec{x}) \ket{\hat{R}_{\pi_{2}}^{bi} \psi_L^0 (-t,\Vec{x})} =
\end{equation*}
\begin{equation*}
     = \bra{\psi_L^0(-t,\Vec{x})} \hat{R}_{\pi_{2}}^{bi \dagger} H_{AV} (-t, \Vec{x})\hat{R}_{\pi_{2}}^{bi} \ket{ \psi_L^0 (-t,\Vec{x})}  =
\end{equation*}
\begin{equation}
   =\bra{\psi_L^0(-t,\Vec{x})} H_{AV} (-t, \Vec{x})\ket{ \psi_L^0 (-t,\Vec{x})},
\end{equation}
where we have used that the Hamiltonian, being a pseudoscalar, is invariant under rotations.
\\
\\
As a partial summary of this section regarding as truly chiral both the system and the influence, we get:
\begin{itemize}
    \item Parity part
\begin{equation*}
    \bra{\psi_L^0 (t,\Vec{x})} H_{AV} (t,\Vec{x})\ket{\psi_L^0 (t,\Vec{x})} =
\end{equation*}
\begin{equation}
    = -\bra{\psi_R^0 (t,-\Vec{x})}  H_{AV} (t,-\Vec{x})\ket{\psi_R^0 (t,-\Vec{x})}.
\end{equation}    
\item Time inversion followed by a $\pi$ rotation
\begin{equation*}
    \bra{\psi_L^0 (t,\Vec{x})} H_{AV} (t,\Vec{x})\ket{\psi_L^0 (t,\Vec{x})} = 
\end{equation*}
\begin{equation}
      =\bra{\psi_L^0(-t,\Vec{x})} H_{AV} (-t, \Vec{x})\ket{ \psi_L^0 (-t,\Vec{x})}.
\end{equation}
\end{itemize}

Hence, apart from the minus sign in the coordinates, the result of this process is the same as in the truly chiral system and influence in the non-QFT case ($H_{LL} = -H_{RR}$).
\\
\\
Regarding this previously mentioned minus sign inside the coordinates, and according to \cite{peskin}, we note that the spatial part of a bispinor can be expressed as
\begin{equation}
    \psi(\Vec{x}) = u^s(p)\exp{-i\Vec{p}\cdot\Vec{x}},
\end{equation}
where
\begin{align}
    u^s(p) & = \begin{pmatrix}
        \sqrt{p\cdot\sigma}\xi^s \\
        \sqrt{p\cdot\Bar{\sigma}}\xi^s
    \end{pmatrix},&
    \xi^1 & = \begin{pmatrix}
        1 \\
        0
    \end{pmatrix},&
        \xi^2 & = \begin{pmatrix}
        0 \\
        1
    \end{pmatrix}.
\end{align}
\\
\\
Thus, a bispinor depending on both spatial and temporal coordinates can be written as
\begin{equation}\label{psiespacialtemporal}
    \psi(\Vec{x},t) =  u^s(p)\exp{-i\Vec{p}\cdot\Vec{x}}\exp{-iEt}.
\end{equation}
\\
\\
Due to the form of our Hamiltonian, \eqref{HEW}, and because we have the same bispinor within both bras and kets, all bispinors are always accompanied by their complex conjugates and, therefore, 
\\
\\
\begin{eqnarray}
    \bra{\psi_L^0}H_{AV}\ket{\psi_L^0} &\overset{\hat{P}}{=}&  -\bra{\psi_R^0}H_{AV}\ket{\psi_R^0} \nonumber \\
    &\overset{\hat{T} \hat{R}_{\pi_{2}}^{bi}}{=}& \bra{\psi_L^0}H_{AV}\ket{\psi_L^0}.
\end{eqnarray}
\\
\\
Therefore, effectively, $H_{LL} = -H_{RR}$ when both the system and the influence are trully chiral and both are treated within a quantum-field-theoretical approach. In this sense, the transformation rules we used for molecular states, specifically regarding time invariance, together with the non-relativistic treatment of the electroweak Hamiltonian, can be considered as nice approximations to the fully rigorous treatment we have performed. 
\\
\\
Although the whole process was performed using a specific Hamiltonian, it can be easily seen that any Hamiltonian that transforms as a truly chiral influence yields the same result (basically, all the signs that come out during the process must be equal to those of the case we have considered). Therefore, under a complete QFT description, we can conclude that a truly chiral influence can induce PVED in molecules.
\\
\\
Let us now go into a specific case of a falsely chiral influence, represented by an an axion-mediating interaction, as pointed out by Barron in \cite{Barron2012}.

\subsection{Axion-mediating interaction in QFT as falsely chiral influence}

As we mentioned in Sec. \ref{sec3}, an axion-mediating interaction between an electron and a nucleon represents a falsely chiral interaction. Before starting with the chirality test,  we have to express the Hamiltonian \eqref{axion}, $H_{ax}$, in a quantum field theoretically way.
\\
\\
We can construct $H_{ax}$ in several ways. Although the electron-axion and nucleon-axion interaction can be generally written as an axial-vector interaction \cite{Di_Luzio_2020}, we must be consistent with the Hamiltonian \eqref{axion}, using scalar and pseudoscalar vertices (see \cite{moody} and \cite{axionesjaja}). With this in mind, we can express the Hamiltonian as

\begin{equation}
H_{ax}(t,\Vec{x}) = -i g_{ea}g_{Na} \Bar{N}(t,\Vec{x})N(t,\Vec{x}) a(t,\Vec{x}) \Bar{e}(t,\Vec{x}) \gamma^5 e(t,\Vec{x}),
\end{equation}
where $g_{ea}$ and $g_{Na}$ are coupling constants, and $a(t, \vec{x})$ is the axion field. It should be noted that the axion field acts like a propagator which, for an axion-type particle (scalar field) is given, in position space, by
\begin{equation}
    \chi_a = \frac{1}{\Box -m^2}.
\end{equation}
After expanding in Taylor series as
\begin{equation}
    \chi_a = \left(-\frac{1}{m^2}\right)\left(1+\frac{\Box}{m^2}+\left(\frac{\Box}{m^2}\right)^2+...\right)
\end{equation}
\\
\\
and assuming low momenta, the first-order propagator becomes

\begin{equation}
    \chi_a = -\frac{1}{m^2}.
\end{equation}
\\
\\
It is noteworthy that this propagator is a time-even scalar, so it is not affected by $\hat{P}$, $\hat{T}$ or $\hat{R}_\theta$ operations. 
\\
\\
For simplicity in the notation, we will join all the coupling constants and the propagator into a single constant, denoted as $K_a$. Thus, the Hamiltonian can be  finally expressed as
\begin{equation}
H_{ax}(t,\Vec{x}) = i K_a \Bar{N}(t,\Vec{x})N(t,\Vec{x})\Bar{e}(t,\Vec{x}) \gamma^5 e(t,\Vec{x}).
\end{equation}
\\
\\
We are now ready to perform the chirality test.

\subsubsection{Behaviour under parity}

The process is similar to the electroweak case, but now we need to use the following transformation rules under parity \cite{peskin}
\begin{equation}
    \hat{P}\Bar{\psi}(t,\Vec{x})\psi(t,\Vec{x})\hat{P} = +\Bar{\psi}(t,-\Vec{x})\psi(t,-\Vec{x}),
\end{equation}
\begin{equation}
    \hat{P}i\Bar{\psi}(t,\Vec{x})\psi(t,\Vec{x})\hat{P} = -i \Bar{\psi}(t,-\Vec{x})\gamma^5\psi(t,-\Vec{x}),
\end{equation}
\\
\\
arriving to
\begin{equation*}
    \bra{\psi_L^0 (t,\Vec{x})} H_{ax} (t,\Vec{x})\ket{\psi_L^0 (t,\Vec{x})} =\bra{\hat{P}\psi_R^0 (t,-\Vec{x})} H_{ax} (t,\Vec{x})\ket{\hat{P}\psi_R^0 (t,-\Vec{x})} =
\end{equation*}
\begin{equation*}
    \bra{\psi_R^0 (t,-\Vec{x})} \hat{P}i K_a\Bar{N}(t,\Vec{x})N(t,\Vec{x})\Bar{e}(t,\Vec{x}) \gamma^5 e(t,\Vec{x})\hat{P}\ket{\psi_R^0 (t,-\Vec{x})} = -\bra{\psi_R^0 (t,-\Vec{x})}  H_{ax} (t,-\Vec{x})\ket{\psi_R^0 (t,-\Vec{x})}.
\end{equation*}

Observe that the characteristic minus sign of the PVED is again obtained. The next step is to apply $\hat{T}$ and $\hat{R}_\pi$, and check if the degeneration of enantiomers can be restored.

\subsubsection{Behaviour under time reversal}
By using the following transformation rules \cite{peskin},
\begin{equation}\label{T3}
    \hat{T}\Bar{\psi}(t,\Vec{x})\psi(t,\Vec{x}) = -\Bar{\psi}(-t,\Vec{x})\psi(-t,\Vec{x})\hat{T},
\end{equation}
\begin{equation}\label{T2}
    \hat{T}i\Bar{\psi}(t,\Vec{x})\gamma^5\psi(t,\Vec{x})\hat{T} = -i \Bar{\psi}(-t,\Vec{x})\gamma^5\psi(-t,\Vec{x}),
\end{equation}
\\
\\
we obtain
\begin{equation*}
     \bra{\psi_L^0 (t,\Vec{x})} H_{ax} (t,\Vec{x})\ket{\psi_L^0 (t,\Vec{x})} = \bra{\hat{T}\phi(-t,\Vec{x})} H_{ax} (t,\Vec{x})\ket{\hat{T}\phi (-t,\Vec{x})} =
\end{equation*}
\begin{equation*}
   = -\left(  \bra{\phi(-t,\Vec{x})}\hat{T}i K_a \Bar{N}(t,\Vec{x})N(t,\Vec{x})\Bar{e}(t,\Vec{x}) \gamma^5 e(t,\Vec{x})\hat{T}\ket{\phi (-t,\Vec{x})} \right)^\dagger=
\end{equation*}
\begin{equation*}
    = -\left(  \bra{\phi(-t,\Vec{x})}i K_a \Bar{N}(-t,\Vec{x})N(-t,\Vec{x})\Bar{e}(-t,\Vec{x}) \gamma^5 e(-t,\Vec{x}) \ket{\phi (-t,\Vec{x})} \right)^\dagger =
\end{equation*}
\begin{equation*}
    = - \bra{\phi(-t,\Vec{x})}\left(i K_a N^\dagger(-t,\Vec{x})\gamma^0 N(-t,\Vec{x})e^\dagger(-t,\Vec{x})\gamma^0 \gamma^5 e(-t,\Vec{x})  \right)^\dagger\ket{\phi (-t,\Vec{x})}=
\end{equation*}
\begin{equation*}
    = - \bra{\phi(-t,\Vec{x})}iK_a \Bar{e}(-t,\Vec{x}) \gamma^5  e(-t,\Vec{x}) \Bar{N}(-t,\Vec{x}) N(-t,\Vec{x})  \ket{\phi (-t,\Vec{x})}=
\end{equation*}
\begin{equation*}
    =- \bra{\phi(-t,\Vec{x})}H_{ax}(-t,\Vec{x}) \ket{\phi (-t,\Vec{x})}.
\end{equation*}

\subsubsection{Behaviour under rotations}
Finally, we can proceed with the rotation part of the chirality test. Using the same reasoning employed in the previous section, we use a $\hat{R}_{\pi_{2}}^{bi}$ rotation. With this in mind, we get
\begin{equation*}
-\bra{\phi(-t,\Vec{x})} H_{ax} (-t, \Vec{x}) \ket{\phi (-t,\Vec{x})}=
\end{equation*}
\begin{equation*}
    = -\bra{\hat{R}_{\pi_{2}}^{bi} \psi_L^0(-t,\Vec{x})} H_{ax} (-t, \Vec{x}) \ket{\hat{R}_{\pi_{2}}^{bi} \psi_L^0 (-t,\Vec{x})} = 
\end{equation*}
\begin{equation*}
     = -\bra{\psi_L^0(-t,\Vec{x})} \hat{R}_{\pi_{2}}^{bi \dagger} H_{ax} (-t, \Vec{x})\hat{R}_{\pi_{2}}^{bi} \ket{ \psi_L^0 (-t,\Vec{x})} =
\end{equation*}
\begin{equation*}
   =-\bra{\psi_L^0(-t,\Vec{x})} H_{ax} (-t, \Vec{x})\ket{ \psi_L^0 (-t,\Vec{x})},
\end{equation*}
noting that the rotation does not affect the Hamiltonian, since it transforms as a pseudoscalar. 
\\
\\
%So, at the end of this whole process (temporal inversion + rotation), the result is
%\begin{equation*}
    %\bra{\psi_L^0 (t,\Vec{x})} H_{ax} (t,\Vec{x})\ket{\psi_L^0 (t,\Vec{x})} =
%\end{equation*}
%\begin{equation}
    %= -\bra{\psi_L^0(-t,\Vec{x})} H_{ax} (-t, \Vec{x})\ket{ \psi_L^0 (-%t,\Vec{x})}
%\end{equation}
Finally, as a partial summary of the whole section, we get
\begin{eqnarray}
    \bra{\psi_L^0} H_{ax}\ket{\psi_L^0} &\overset{\hat{P}}{=}&  -\bra{\psi_R^0}  H_{ax}\ket{\psi_R^0}  \nonumber \\
     &\overset{\hat{T}\hat{R}_{\pi_{2}}^{bi}}{=}& -\bra{\psi_L^0} H_{ax}\ket{ \psi_L^0}.    
\end{eqnarray}

Therefore, our result is compatible with the ii) case of the Sec. \ref{sec3} referring to a falsely chiral influence acting on molecules, where we reached $H_{LL} = H_{RR} = 0$. In this sense, the QFT approach reinforces both the conditions \eqref{tqcond} and the non-relativistic approximation of the electron-nucleon interaction mediated by an axion as good approximations, and it shows that a false chiral influence can not induce PVED in molecules.

\section{Final comments and conclusions}\label{sec5}

\begin{center}
    
    \begin{figure}[ht]
\centering
\includegraphics[width= 13.5cm]{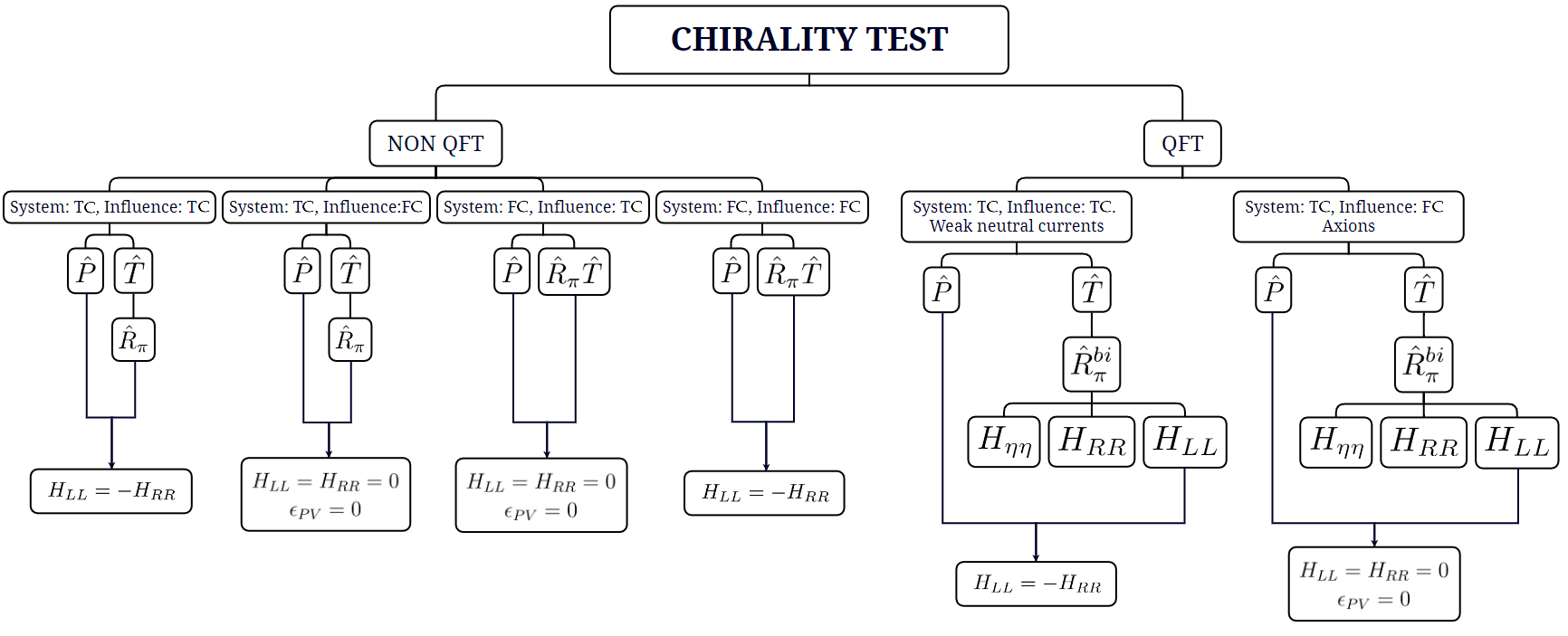}

\caption{\justifying Description of the "chirality test'', which consists in applying the parity, $\hat P$, time inversion, $\hat T$, and a $\pi$ rotation operator, $\hat R_{\pi}$, to both systems and influences, which can be taken to be truly or falsely chiral. Regarding the application of the test to the QFT part, we remind the reader that $\hat R^{bi}_{\pi}$ refers to a rotation operator for bispinors which can be left (L)-, right (R)-handed or neither of them (denoted by $\eta$). See text for details.}
\label{testfig}
\end{figure}
\end{center}

The basic idea of this manuscript has been to apply what we called the "chirality test", which Fig. (\ref{testfig}) summarizes.

In essence, in the first part of the manuscript, which is referred to as "non QFT" in  Fig. (\ref{testfig}), this test consists of applying the parity, $\hat P$, time inversion, $\hat T$, and a $\pi$ rotation operator, $\hat R_{\pi}$, to both systems and influences, which can be taken to be truly or falsely chiral.
\\
\\
Specifically, we have explored which type of interactions can lift the degeneracy between truly and falsely chiral systems, showing that
only when both systems and influences are both truly (falsely) chiral, a parity violating energy difference between left- and right-handed systems can be produced. 
\\
\\
Along the second part, denoted by ``QFT" in Fig. (\ref{testfig}), we have established the following facts: (i) on one hand, after considering the enantiomers of a chiral molecule as paradigmatic truly chiral systems, we rigorously showed, under a quantum field theoretically approach, that only a truly chiral influence such as the $Z^{0}$-mediated electroweak interaction can lift the degeneracy between enantiomers. In this sense, Barron's result \cite{barronpv} regarding that only a truly chiral influence can induce PVED has been rigorously derived
within quantum field theory; (ii) on the other hand, we have explicitly shown, under the same theoretical framework, that a falsely chiral influence, such as an axion-mediated interaction in chiral molecules, can not lift the aforementioned degeneracy. 
\\
\\
Therefore, we can conclude that Barron’s seminal results concerning true and false chirality remain valid under a quantum field theory-based approach.  We think that this is of importance for a rigorous understanding of a fundamental physics basis of molecular homochirality.

\section*{Acknowledgements}
Discussions with several members of the Relativistic Astrophysics Group (Universidad de Alicante) are gratefully acknowledged.
D. M. -G. acknowledges Fundación Humanismo y Ciencia for financial support. P. B. acknowledges financial support from the Generalitat Valenciana through PROMETEO PROJECT CIPROM/2022/13. 

\bibliographystyle{unsrt}
\bibliography{referencias.bib}

\end{document}